# Ultrasensitive Electrochemical Sensor for Perfluorooctanoic Acid Detection Using Two-dimensional Aluminium Quasicrystal

*Anyesha Chakraborty[a], Raphael Tromer[b], Thakur Prasad Yadav[c], Nilay Krishna Mukhopadhyay[d], Basudev Lahiri[e], Rahul Rao, Ajit.K.Roy, Nirupam Aich, Cristiano F. Woellner, Douglas S. Galvao[b*], Chandra Sekhar Tiwary[g]\**

[a] School of Nano Science and Technology, Indian Institute of Technology, Kharagpur, Kharagpur 721302, India

[b] Applied Physics Department, and Center for Computational Engineering & Sciences (CCES), State University of Campinas, Campinas, SP 13083-970, Brazil

[c] Department of Physics, Faculty of Science, University of Allahabad, Prayagraj 211002 UP, India

[d] Department of Metallurgical Engineering, Indian Institute of Technology (BHU), Varanasi 221005, UP, India

[e] Department of Electronics and Electrical Communication Engineering, Indian Institute of Technology Kharagpur, Kharagpur 721302, India

[f] Department of Metallurgical and Materials Engineering, Indian Institute of Technology Kharagpur, Kharagpur 721302, India

**Abstract**

Per- and polyfluoroalkyl substances (PFAS), often referred as "forever chemicals," are pervasive environmental pollutants due to their resistance to degradation. Among these, perfluorooctanoic acid (PFOA) poses significant threats to human health, contaminating water sources globally. Here, we have demonstrated the potential of a novel electrochemical sensor based on two-dimensional (2D) aluminium-based multicomponent quasicrystals (2D-Al QC) for the ultrasensitive sub-picomolar level detection of PFOA. The 2D-Al QC-inked electrode was employed here to detect PFOA by differential pulse voltammetry (DPV). The limit of detection (LoD) achieved is 0.59± 0.05 pM. The sensor was evaluated for selectivity with other interfering compounds, repeatability of cycles, and reproducibility for five similar electrodes with a deviation of 0.8 %. The stability of the sensor has also been analysed after ninety days ,which shows a minimal variation of 15%. Spectroscopic techniques and theoretical calculations were further utilized to understand the interaction between the 2D-Al QC and PFOA. The results demonstrate that the 2D-Al QC offers a promising platform for the rapid and sensitive detection of PFOA, potentially addressing current environmental monitoring challenges.

**Introduction**

Per- and polyfluoroalkyl substances (PFAS) are a group of compounds that are currently creating an increased threat to environment and human health.[1] These ubiquitous compounds are commonly used in products like fire retardants, stain repellents, non-stick cookware, oil and gas industry, mining, textiles etc.[2] Generally, PFAS are resistant to degradation even in high temperature, extreme pressure and highly-corrosive environment,[3] due to the presence of the aliphatic fluorocarbon (C-F) bonds in their chemical formula, for which they are also known as 'forever chemicals'.[2] They are widely used for their structural resilience, conversely this nature renders them as a menace to the environment. The high fluorinated group makes them hydrophobic and lipophobic, though the functional group helps them to interact with polar medium.[3] Exposure to PFAS can pose several health risks to liver, thyroid, immune system and reproductive systems, besides increased cholesterol, obesity, cancer.[1,4] PFOA is a widely used compounds of PFAS family. Being water soluble, it readily dissolves in ground water, leading to water contamination. According to United States Environmental Protection Act 2016, the proposed maximum contaminant level of PFOA in drinking water was 70 ng/L[2] which was further lowered in 2022 to 4 ng/L.[5] Currently, PFOA sensing by an ultrasensitive system is very crucial at the onset of the pollution, for prevention and monitoring . To trace this compound various detection methods has been employed. Conventional detection methods like chromatography, and mass spectrometry though utilised as standard methods of detection, are not practical as rapid detection technique for regular monitoring.[6] Recently researchers have utilised optical detection methods like fluorescence spectroscopy[5,7] conjugated with other methods for PFAS sensing such as employing fluorescent polymers,[5] glass-fabricated chip,[7] utilising luminescent metal organic framework sensor array,[8] ssDNA aptamer,[9] and fluorescent imprint-and report sensor array.[10] Again, some approaches has been taken like utilization of liquid interfacial properties,[11] energy transfer mechanism,[12] for PFAS detection. Electrochemical sensing is a facile, robust, portable and efficient technique for highly specific and ultrasensitive detection of PFAS from soil and water. Recently, electrochemical sensing of various PFAS molecules has conducted using perfluorodecanethiol infused gold nanoparticles,[13] molecularly imprinted polymer (MIP),[14,15] gold electrode modified by MIP made by electropolymerization of o-phenylenediamine,[16,17] silver nanoparticles,[18] MXene-decorated silver nanoparticles.[19] Here, a two-dimensional (2D) aluminium based multicomponent quasicrystals (2D-$Al_{70}Co_{10}Fe_5Ni_{10}Cu_5$ (2D-Al QC)) has been utilised for electrochemical sensing of PFOA. Quasicrystals are regarded as a novel category of intermetallic compounds characterized by their long-range quasiperiodic arrangement.[20] The combination of strong intra-layer covalent bonding and weak inter-layer bonding facilitates them to form 2D structure readily using exfoliation technique.[21] The large surface area and abundant multiple active sites of the material allows better binding of the analytes. They also exhibit high electron mobility, unique surface chemistry, tunable optical and electronic properties, and excellent environmental stability [12] making them an attractive platform for sensors.[21] The monolayered quasicrystals have been widely utilised for catalysis,[22] hydrogen evolution,[21,23] gas sensing,[24,25] and non-linear optics.[26,27]

In the current study, we have synthesised the 2D-Al QC through a highly scalable liquid exfoliation technique, starting from its bulk counterpart. After analysing the structural characteristics of the 2D material thoroughly, the resulting material has been dropcasted onto a glassy carbon electrode (GCE) to create a 2D-Al QC-inked electrode. This electrode has been utilised as the working electrode for the successful detection of PFOA by differential pulse voltammetry (DPV). An analysis has been conducted to evaluate the selectivity, repeatability, reproducibility, and stability of the sensor. Finally spectroscopic techniques and theoretical calculations have been utilised to investigate the interaction between the 2D-Al QC and PFOA thoroughly for better understanding.

**Experimental details**

**Materials and reagents.**

Aluminium (99.5 % purity, Alfa Aesar), Cobalt (99.9 % purity, Alfa Aesar) Nickel (99.5 % purity, Alfa Aesar) Iron (99.9 % purity, Alfa Aesar) Copper (99.5 % purity, Alfa Aesar) Perfluorooctanoic acid (95 %, Sigma Aldrich), $K_4[Fe(CN)_6]$ (98%, Loba Chemie), urea (99 %, Loba chemie), lactic acid (99.5 %, Loba chemie), ammonium chloride (99 %, Loba chemie), Ciprofloxacin (99.5 %, Merck), Roxithromycin (99.5 %, Merck), Ampicillin (99.5 %, Merck), Methylene blue (99.5 %, Loba Chemie), Rhodamine D (99.5 %, Loba Chemie) Isopropyl alcohol (IPA) (99.9 % pure, Merck), Deionised water (DI water) Millipore.

**Material Synthesis and Characterization.**

A polycrystalline alloy has been synthesized by melting Al, Co, Fe, Ni, and Cu metals with high purity using arc melting process, in a stoichiometric proportion of $Al_{70}Co_{10}Fe_5Ni_{10}Cu_5$. Prior to melting, the chamber was purified using a titanium (Ti) button, due to its oxygen-absorbing properties. The alloy ingot was then sealed in a quartz tube and annealed in a furnace at 1000°C for 48 hours under an argon (Ar) atmosphere. The ingot was brittle enough to be crushed into a fine powder. After that, 1 gram of the fine powder was taken in a beaker with 100 ml isopropyl alcohol (IPA) solvent and sonicated it for 5 hours with 30kHz frequency, utilising ultrasonic probe sonicator (Rivotek) at room temperature. Later the suspension was kept at rest for 14 hours and the heavy particles sedimented at the bottom of the beaker. The lightweighted 2D-flakes floated in the solvent. This solvent containing 2D-sheets has been separated gently by micropipette and utilised in further characterisation and application.

The phase identification of the 2D-Al QC carried out using X-ray diffraction (XRD) was recorded by an X-ray diffractometer (*Bruker D8 Advance*) with Cu Kα radiation of wavelength λ = 0.15406 nm. The elemental composition has been determined by the Energy-dispersive spectroscopy (EDS) spectra and the morphology of the semi-bulk 2D-Al QC and the 2D-Al QC coated electrodes both before and after utilisation for sensing, were obtained using Field Emission Scanning electron microscope *(Zeiss Merlin with GEMINI II column)* with operating voltage 5kV. The atomic orientation of the 2D flakes was investigated in Scanning Transmission Electron Microscope-High Angle Annular Dark Field (STEM-HAADF) (*Jeol, F30, FEI*) operating at 300 kV. The thickness of the 2D flakes was measured in the Atomic Force Microscope (AFM) (*Agilent Technologies model no. 5500*) in tapping mode with $Si_3N_4$ tip. All the Raman spectroscopic study and Raman mapping were done in WITec UHTS Raman spectrometer (*WITec, UHTS 300 VIS, Germany*) with a 532 nm laser source and 17 mW laser power, using 1800 lines/mm grating in 2 sec integration time and 10 accumulations for each data. The sample has been prepared for Raman by dropcasting the 2D material, the analyte or the mixture of 2D-material and analyte on a 1 cm×1 cm $Si/SiO_2$ wafer. The FTIR study was done by FTIR spectrometer, *NICOLET 6700* model, *(Thermofisher Scientific Instruments)* in transmission mode dropcasting on KBr pellets.

**Preparation of modified electrodes and electrochemical measurements.**

Initially, the 3 mm GCE was carefully polished using 0.3 microns and 0.05 microns alumina powder and thoroughly cleaned with deionized water (DI water) and ethanol, alternatively. Then, by ultrasonication it has been ensured that even the tiny particles of contaminants were removed from the surface of the electrode. Afterwards, the GCE was coated with the freshly prepared 2D-Al QC ink solution using the dropcasting method. To prepare the ink, a solution of polyvinylidene fluoride (PVDF) of 10mg/ml concentration in acetone solvent was prepared beforehand, which acts as a binder here. An amount of 200 µL of the binder was added to the

5 ml of liquid-exfoliated 2D-Al QC. The mixture was subjected to ultrasonication for 15 minutes and subsequently stirred at room temperature using magnetic stirrer to achieve homogenity. Then this solution kept stable for a duration of 12 hours. Afterwards, the settled gel like material was obtained as ink and applied over the cleansed GCE by dropcasting, followed by drying completely.

The electrochemical measurements were conducted utilizing a Squidstat Plus potentiostat (manufactured by Admiral instruments) with a three-electrode cell setup. In this setup, there was a platinum (Pt) wire used as the counter electrode, Ag/AgCl was employed as the reference electrode, and either a bare glassy carbon electrode (GCE) or a GCE coated with 2D-Al QC ink was used as the working electrode. Cyclic Voltammetry (CV) measurements were conducted within the voltage range of -0.5 to +1.2 V, using a scan rate of 20 mV.s$^{-1}$. DPV measurements were conducted within a scan range of 0 to 1.2 V, using a pulse amplitude of 10 mV, pulse width of 50 ms, and pulse period of 200 ms. Electrochemical impedance spectroscopy (EIS) measurements were performed within a frequency range of 0.1 MHz to 0.1 Hz, with an AC excitation amplitude of 10 mV. In order to examine the electrochemical characteristics of the 2D-Al QC-inked electrode compared to the bare GCE, CV and EIS were conducted in a solution containing 5 mM [Fe(CN)$_6$]$^{3-/4-}$ and 0.1 M KCl.

**Computational details**

Density Functional Theory (DFT) simulations were carried out using the SIESTA code [ref] at the local density approximation (LDA) for the exchange-correlation. SIESTA employs a linear combination of numerical atomic orbitals. In this study, a double-ζ basis set with polarization functions (DZP) was used. The interactions between electrons and core ions were modeled with the norm-conserving Troullier-Martins pseudopotential [ref] in its Kleinman-Bylander form [ref]. An energy mesh cutoff of 250 Ry was set to ensure accurate charge density representation in real space. Since the investigated were finite, they were treated as large "molecules," and calculations were performed only at the gamma point of the special k-point mesh grid. The SCF convergence criteria required the maximum difference between input and output density matrix elements to be less than $10^{-1}$ eV. For ionic optimization, convergence was achieved when atomic forces were less than 0.05 eV/Å. We used a rectangular 'supercell', with the initial configuration having atoms and molecules randomly distributed across the QC layer.

To investigate the interaction between the 2D-Al QC and the PFOA molecule, we initially randomly placed the molecule at different surface positions and carried out geometry optimization for each configuration. For each optimized position, we subsequently carried SIESTA ab initio molecular dynamics (AIMD) simulations, considering room temperature in an NVT ensemble for a total simulation time of 20 ps with time steps of XX.

# Results and Discussion

## Structural Characteristics of 2D-Al QC

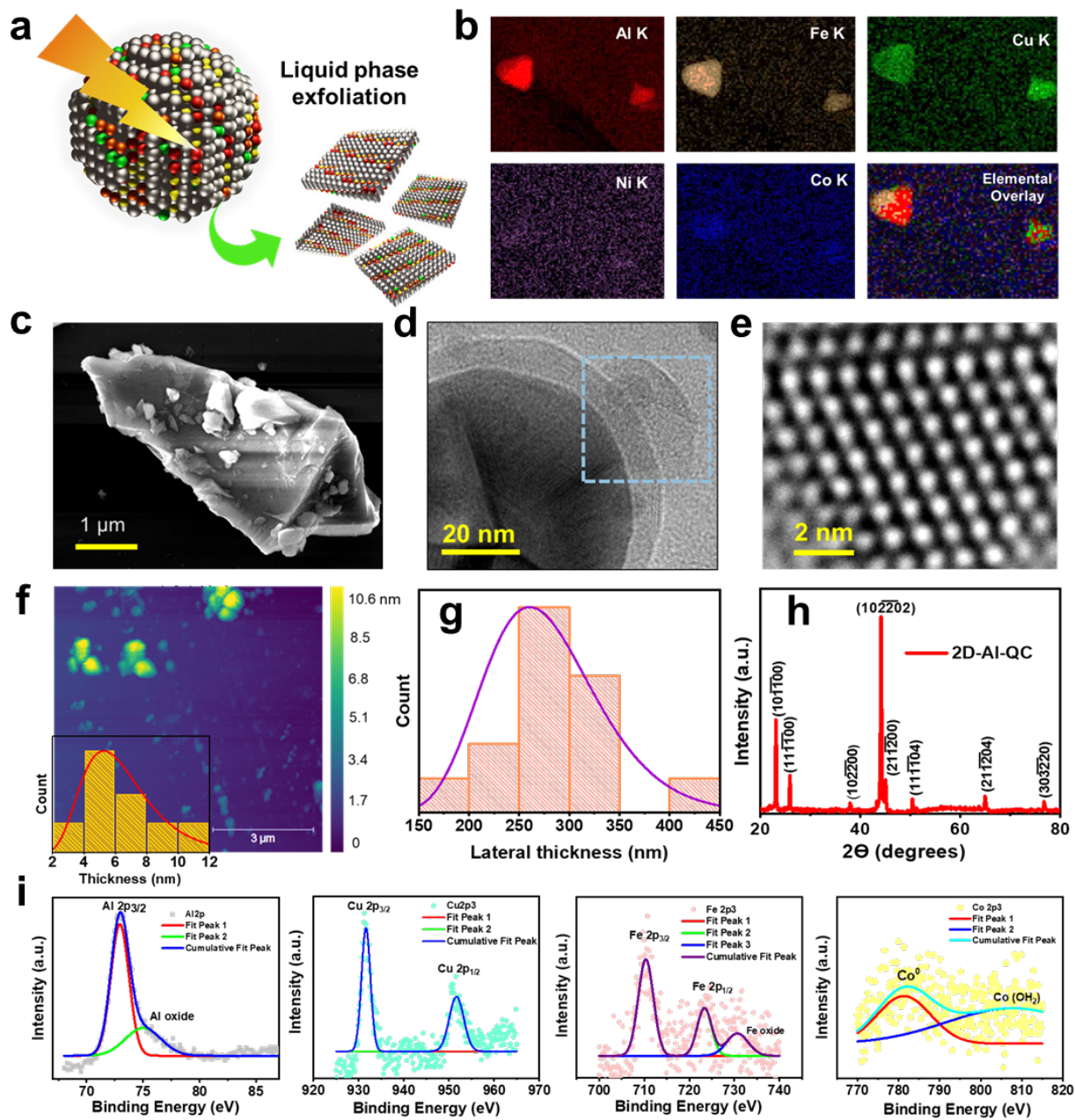

*Figure 1.* a) Schematic representation of the synthesis of 2D-sheets from bulk, b) EDS mapping of the exfoliated 2D- Al QC, c) FESEM image of semi-bulk Al-QC after 2hr exfoliation, d)

*HRTEM image of exfoliated 2D-sheets, e) High resolution HRTEM image of the 2D-sheet, f) AFM image of 2D-flakes with histogram of thickness profile in the inset, g) Lateral thickness histogram plot, h) XRD pattern of 2D-Al QC, i) High resolution XPS of the elements of 2D-Al QC.*

**Figure 1a** represents the schematic diagram of the synthesis of the multicomponent 2D- Al QC sheets from the bulk $Al_{70}Co_{10}Fe_5Ni_{10}Cu_5$ by liquid phase exfoliation utilising IPA as solvent. The detail process has been explained in the Experimental section. **Figure 1b** depicts the elemental mapping of all the elements present in 2D-Al QC and the elemental overlay image obtained by Energy Dispersive X-ray (EDX). This shows the elements are in following proportion in the 2D-Al QC sheets: Al is 67 %, Cu 13%, Fe 11%, Co 5 %, and Ni 4 %.

The SEM image of the semi-bulk flake of the multielement Al QC, after 2 hours of exfoliation is present in **Figure 1c.** This image shows the brittle nature of the material, which helps in fragmentation and thinning of the material upto the formation of 2D sheets with further exfoliation. The EDS spectra of **Figure S1** confirming the elements with it's atomic percentage present in the table. This signifies that the bulk QC have the accurate molar ratio which has been further changed while 2D formation during exfoliation.[21,28] The elements like Cu and Fe which were initially of 5% each in the bulk form, get exposed on the surface in 2D flakes due to exfoliation. It leads to enhancement of multiple active sites in the material which favours in easy binding with analyte, thus enhance sensitivity. Next, the liquid exfoliated 2D-Al QC has been structurally characterised using HRTEM in **Figure 1d.** To obtain the results, the 10 µl of IPA-exfoliated 2D-Al QC has been dropcasted on the TEM grid and dried properly. **Figure 1d** shows the HRTEM image of the layer-by-layer exfoliation of atomically thin 2D-flakes from its bulk counterpart. In **Figure 1e**, the high resolution image of HRTEM has been displayed which shows the atomic arrangements of the 2D sheet. The thickness of the 2D flakes can be determined by the AFM image present in **Figure 1f.** The histogram plot of thickness represents that the thickness of all the flakes is between 2-12 nm, which confirms the formation of two-dimensional system. Moreover, maximum flakes are in 4-8 nm range, which ensures the few atomic layered thickness of the 2D-flakes. The lateral thickness histogram present in **Figure 1g** shows the lateral dimension of all the flakes ranges from 150 nm to 450 nm. Maximum number of flakes are within 250-350 nm range. The XRD pattern present in **Figure 1h** perfectly matches with the Al-based decagonal quasi crystal,[23] thus confirms the formation of 2D-Al QC. **Figure 1i** defines the chemical composition of the 2D-Al QC by high resolution XPS analysis. **Figure 1i (i)** shows the high resolution XPS spectra of Al 2p. The intense peak at 73 eV denotes to Al $2p_{3/2}$ with a minor oxide peak at 75 eV which generates due to exfoliation.[23] Next **Figure 1i (ii)** displays the high resolution XPS of Cu 2p having the intense peaks at 932 eV and 952 eV confirming the Cu $2p_{3/2}$ and Cu $2p_{1/2}$

**Electrochemical properties of 2D-Al QC**

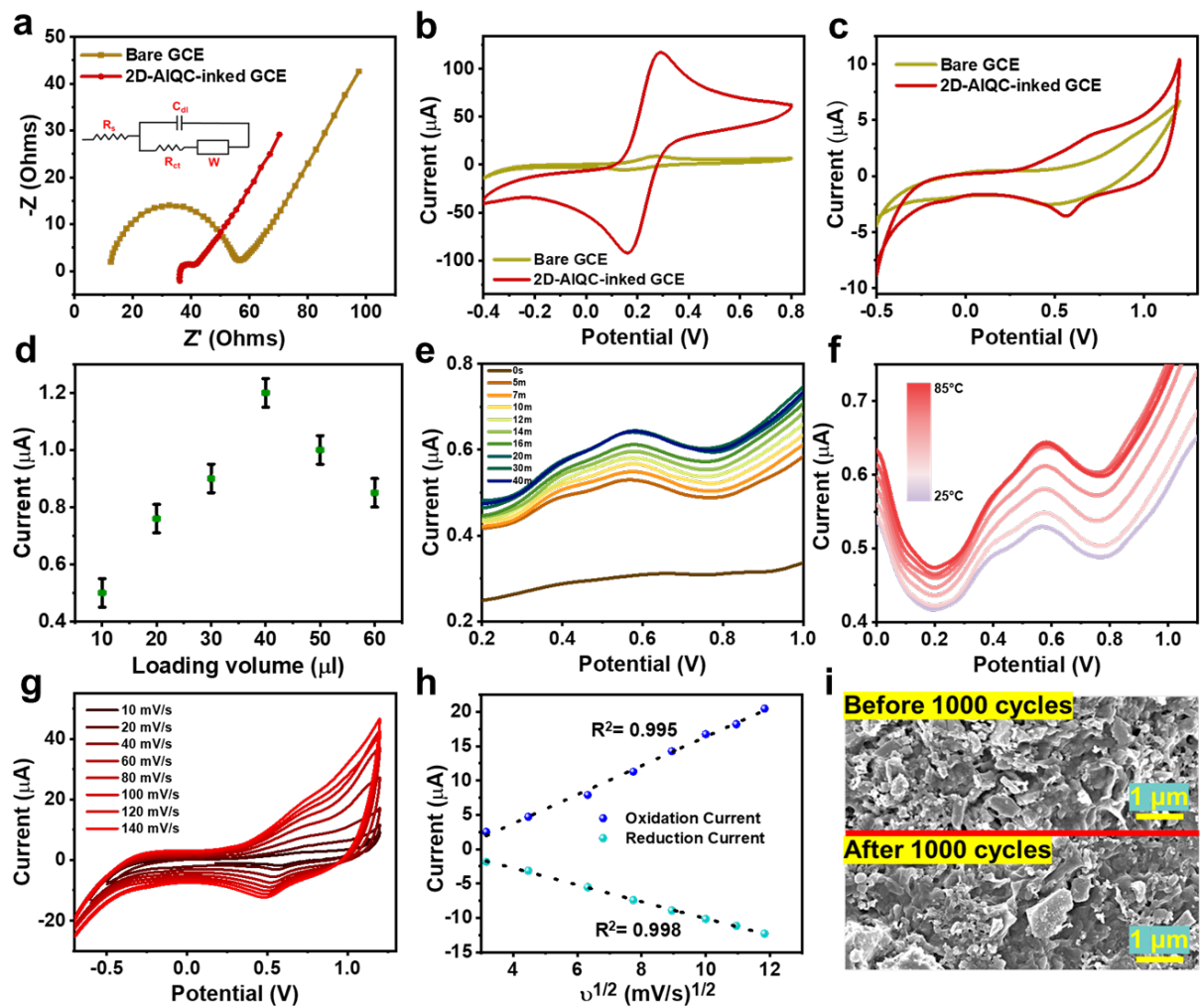

*Figure 3.* a) Nyquist plots of bare GCE and 2D-AlQC-inked GCE in 5 mM [Fe(CN)$_6$]$^{3-/4-}$ with 0.1 M KCl and the Randle's circuit diagram in the inset, b) CV curves of bare GCE and 2D-AlQC-inked GCE in 5 mM [Fe(CN)$_6$]$^{3-/4-}$ with 0.1 M KCl, c) CV curves of bare GCE and 2D-AlQC-inked GCE in 1 µM PFOA dissolves in 100 ml deionised water, d) Effect of loading volume of 2D-AlQC on the oxidation current in 1 µM PFOA solution, e) Effect of incubation time on the oxidation current in 1 pM PFOA solution, f) Temperature dependent DPV curves in 1 pM PFOA solution, g) Scan rate variation, g) Calibration curve demonstrating the variation of oxidation and reduction current with square root of the scan rates, h) SEM images of 2D-AlQC-inked GCE before and after 1000 cycles of CV in 1 µM PFOA solution.

The electrochemical characteristics of the 2D-Al QC-coated GCE were analyzed using EIS and CV in 5 mM Fe(CN)$_6^{3-/4-}$ and 0.1 M KCl, as shown in **Figures 3a and 3b**, respectively. In **Figure 3a**, the Nyquist plots for both the bare GCE and 2D-Al QC-coated GCE are presented, where the semicircle at the high-frequency region indicates the electron transfer limiting process. The diameter of this semicircle signifies the charge transfer resistance ($R_{ct}$), while the linear section in the low-frequency region corresponds to the diffusion-controlled process.[29] The larger semicircle diameter observed for the bare GCE signifies a higher charge transfer resistance, whereas the 2D-Al QC-coated GCE exhibits a much smaller semicircle, indicating lower charge transfer resistance and enhanced conductivity in electrochemical reactions. The Randles circuit, displayed in the inset of **Figure 3a**, used to fit the impedance data, where $R_s$, $R_{ct}$, $C_{dl}$, and W represent solution resistance, charge transfer resistance, double-layer capacitance, and Warburg impedance, respectively. The charge transfer resistance for the bare GCE was found to be 38.5 Ω, while for the 2D-Al QC-coated electrode, it was 3.95 Ω. **Table S1** summarizes the values of all the elements derived from the Randles circuit.

To evaluate the electrochemically active surface area of the 2D-Al QC-coated GCE, CV data were collected for both bare GCE and 2D-Al QC-coated GCE in 5 mM [Fe(CN)$_6$]$^{3-/4-}$ with 0.1 M KCl at a scan rate of 20 mV/s, as shown in **Figure 3b**. The 2D-Al QC-coated GCE produced oxidation and reduction currents of +116.7 μA and -91.8 μA, respectively, with a peak potential difference (ΔEp) of 0.12 V. In contrast, the bare GCE exhibited oxidation and reduction currents of 8.65 μA and -5.68 μA, respectively, with a ΔEp of 0.172 V. The higher redox currents and smaller peak potential separation of the modified electrode indicate faster electron transfer kinetics.[30] The effective active surface area of the electrodes was calculated using the Randles-Servick equation (Equation 1) as follows[30]:

$$I_p = 2.69 \times 10^5 \, AD^{1/2} \, n^{3/2} \, \gamma^{1/2} C \tag{1}$$

where $I_p$ is the peak current (A), A is the active surface area (cm²), D is signified as the diffusion coefficient of $[Fe(CN)_6]^{3-/4-}$ (cm².s⁻¹) here, D = 7.6 × 10⁻⁶ cm²/s, n denotes the number of electron transfers, γ is represented as applied scan rate (Vs⁻¹) and, C is the concentration of $[Fe(CN)_6]^{3-/4-}$ (mol.cm⁻³) in solution. The calculated effective surface areas for the bare GCE and the 2D-Al QC-coated GCE were 0.073 cm² and 0.223 cm², respectively.

The electrochemical performance of the 2D-Al QC-coated GCE in PFOA detection was further explored by comparing it to the bare GCE in 1 µM PFOA in deionized water at a scan rate of 20 mV/s in **Figure 3c**. The 2D-Al QC-coated electrode exhibited an oxidation peak current of 3.78 µA at 0.72 V and a reduction peak current of -3.82 µA at 0.56 V, while the bare GCE displayed an oxidation current of 2.58 µA at 0.93 V and a reduction current of -2.5 µA at 0.47 V. The higher redox currents of the modified electrode highlight its better electrochemical performance in detecting PFOA.

The effect of the loading volume of 2D-AlQC on response peak current has been evaluated in **Figure 3d** by varying the loading amount from 10 µl to 60 µl in 1 µM PFOA. With increasing amounts, the peak current increases from 10 to 40 µl due to the enhancement of 2D-AlQC on the electrode surface, and after 40 µl current decreases due to the thickening of 2D-flakes on the electrode surface. The effect of incubation time is shown in **Figure 3e.** With increasing time from 0 s to 20 min the oxidation current of DPV increases, after 20 min it gets stagnant showing the maximum current at 20 min incubation time. In **Figure 3f** the temperature-dependent DPV has been analysed with varying temperature from 25 to 85 °C. The oxidation current increases from 25 °C to 65 °C and remains constant after that, which depicts the thermal stability of the 2D-AlQC. The effect of scan rate variation has been analysed in **Figure 3g** with 1 µM PFOA in DI water with variation of the sweep rate from 10 mV/s to 140 mV/s. The calibration curve plotted in **Figure 3h** shows the oxidation and reduction peak currents increased linearly with the square root of scan rates with linear correlation coefficient values

($R^2$) 0.995 and 0.998 respectively. This confirms that the oxidation and reduction of PFOA on a 2D-AlQC-coated electrode is a diffusion-controlled process.[31] The morphology and structure of the 2D-AlQC-inked electrode have been investigated before and after electrochemical experiments using FESEM images in **Figure 3i.** The morphological stability of the 2D-AlQC-inked electrode has been confirmed by comparing the FESEM images of the electrode before testing and after 1000 cycles of cyclic voltammetry.

**Electrochemical Sensing of PFOA**

The electrochemical sensing of PFOA has been done by DPV with varying concentrations of PFOA from 1 µM to 1 pM. From **Figure 4a**, it can be observed that with increasing PFOA concentration, the current increases. The calibration curve obtained in **Figure 4b** shows the linear relationship of the current with PFOA concentration represented in a logarithmic scale with regression coefficient $R^2 = 0.945$. The LoD calculated from the calibration curve is 0.59 ± 0.03 pM. Limit of detection has been calculated by using the following formula,[32]

$$\text{Limit of Detection} = 3.3 \times Standard\ Deviation/Slope \tag{4}$$

The selectivity of the 2D-AlQC inked electrode has been checked in **Figure 4c** using 10 mM ciprofloxacin, 10 mM urea, 10 mM lactic acid (LA), 10 mM $NH_4Cl$, 10 mM methylene blue, 10 mM ampicillin, 10mM rhodamine D dye and 10 mM roxithromycin, all the closely interfering organic compounds of PFOA that might be present in water. It is evident from the DPV curves in **Figure 4c** that only PFOA exhibits a pronounced peak, and ciprofloxacin shows a minor peak. The prominent identification of PFOA in the mixture confirms the high selectivity and anti-interfering nature of the sensor. In **Figure 4d** the stability or longevity of the sensor has been assessed for 3 months, after every 30 days, starting from the 1st day of operation. In the whole process the inked electrode kept in an air-tight container at room temperature to avoid the contact of moisture. After the first 30 days the sensor results slightly decreased by 3.33% and after next 30 days *i.e.* 60 days from the 1st day of operation it decreased by 10 %. Finally, after 90 days the current response reduces by 15% probably due to the reduction of active sites and oxidation of the 2D flakes. To check the repeatability, we have studied 20 cycles of CV in the same electrode, and the result is presented in **Figure 4e** depicting the excellent repeatability of the sensor. To study the reproducibility, five different inked electrodes have been prepared with identical 2D-Al QC ink formulation. The result presented in **Figure 4f** shows excellent reproducibility with at most 0.8 % variation.

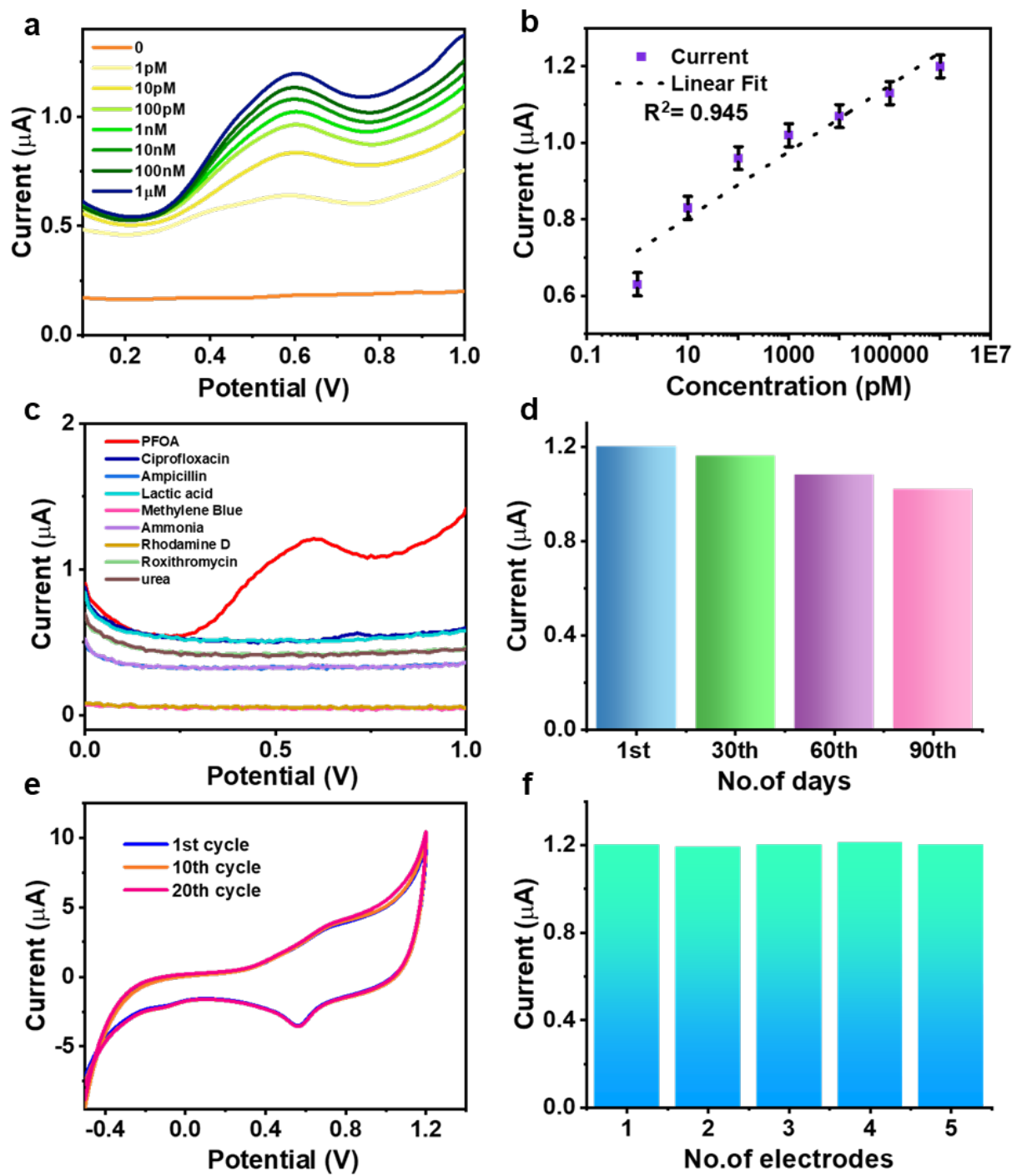

**Figure 4.** *a) DPV curves of 2D-AlQC-inked electrode with varying concentration of PFOA, b) Calibration curve of varying concentration of PFOA, c) Selectivity test, d) Stability analysis for 3 months after every 30 days, e) Repeatability test, and f) Reproducibility analysis with five different modified working electrodes.*

**Spectroscopic investigations of the interaction of 2D-AlQC and PFOA**

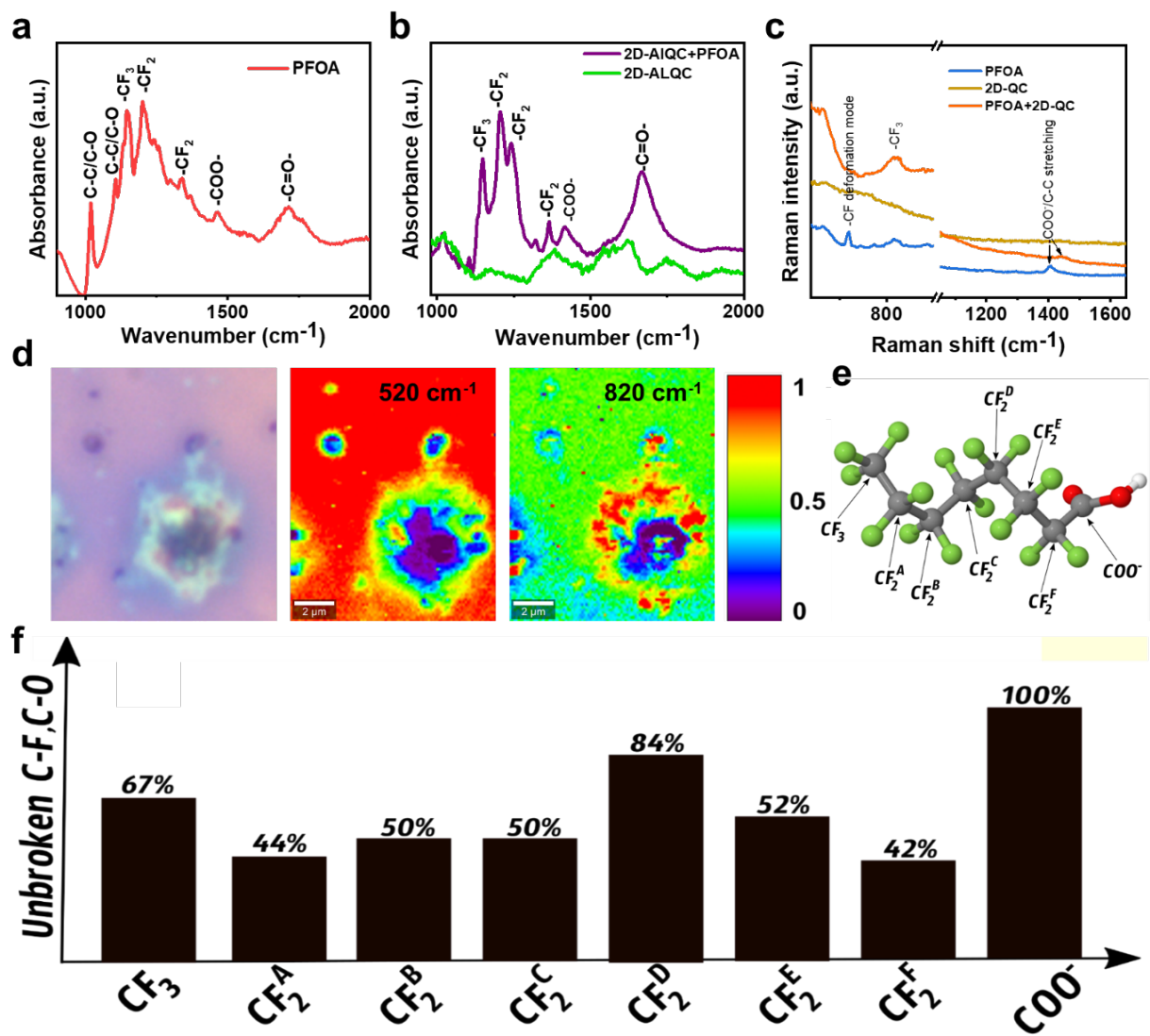

**Figure 5**.*a) FTIR spectra of PFOA, b) FTIR spectra of 2D-QC and the mixture of PFOA with 2D-QC, c) Raman spectra of PFOA, 2D-QC and the mixture of the PFOA with 2D-QC, d) Optical image of the 2D-QC-PFOA mixture and Raman images. e) PFOA molecule indicating its functional groups, and f) percentual of the remaining unbroken C or O with F bonds.*

The interaction of PFOA with the 2D-Al QC has been studied thoroughly by spectroscopic analysis. In **Figure 5a** the IR spectra of PFOA alone is presented. The most intense absorption peaks at 1145 cm$^{-1}$, 1200 cm$^{-1}$ and 1345 cm$^{-1}$ is designated to -$CF_3$ and -$CF_2$ respectively[33] and the other peaks at 1019 cm$^{-1}$ and 1105 cm$^{-1}$ indicates the C-C and C-O stretching, [34,35] and 1465

cm$^{-1}$ and 1714 cm$^{-1}$ represents -COO- asymmetrical stretching and -C=O- stretching.[36] After the interaction of PFOA with 2D-Al QC in **Figure 5b** it is shown that the prominent -CF$_2$ and -CF$_3$ peaks are shifted towards higher wavenumber at 1148 cm$^{-1}$, 1205 cm$^{-1}$, 1245 cm$^{-1}$ and 1364 cm$^{-1}$ respectively indicating the adsorption of the -CF$_2$ and -CF$_3$ bonds to the 2D-Al QC. Moreover, the intensity of the peaks depicting -C-C/C-O- bonds has been diminished largely, though the -COO- and -C=O- remains intact which has been completely co-related with the fact presented later in theoretical part. **Figure 5c** represents the Raman spectra of the PFOA alone, 2D-Al QC and the mixture of PFOA and 2D-Al QC to compliment the IR study with the Raman spectra analysis and Raman mapping. The Raman peak at 820 cm$^{-1}$ corresponds to -CF bending has been vividly observed in the mixture of the PFOA and 2D-Al QC which depicts the interaction of the PFOA with 2D-Al QC.[37] In **Figure 5d** the optical microscopic image of the 2D-Al QC and PFOA mixture at 100X magnification. The sample has been prepared by dropcasting the mixture on a silicon substrate. Next, the Raman mapping has been obtained in that dropcasted sample at that particular area in 0.05 s integration time at 100 X magnification. The Raman maps of the particular region at different wavenumbers denote the presence of the 2D-Al QC and PFOA by their respective vibrational modes. The red coloured portion denotes the highest intensity, the purple the lowest and the green the middle intensity in the Raman maps. At 520 cm$^{-1}$, the red-coloured portion indicates the silicon substrate and the purple to-blue colour denotes the 2D material and the green to yellow portion indicates the PFOA. Similarly, in 820 cm$^{-1}$, the PFOA portion is in red, as PFOA has an elevated peak at that wavenumber due to C-F bending and the 2D-material portion is in purple to blue. This indicates the interaction of the 2D material with PFOA molecules.

**Theoretical Analysis of the Interaction of PFOA with 2D-QC**

In the experimental section, we discussed the electrochemical details of how the interaction between PFOA and the 2D-Al QC crystal can be mapped with respect to the analyses performed on the isolated parts of each component. By analyzing the results, especially the Raman results shown in Figure 5, we can conclude that the interaction between PFOA and the 2D-Al QC induces PFOA significant structural changes. Some normal vibrational modes corresponding to the $CF_3$ and $COO^-$ functional groups located at the PFOA chain ends remain after the interaction, while the $CF_2$ groups (C-F bond) located throughout the molecular chain tend to dissociate.

To gain further insights into interpreting the experimental data, we carried out *ab initio* molecular dynamics (AIMD) considering the system (PFOA + 2D-Al QC) in a fixed volume at room temperature (300 K) for 1.9 ps. We considered six cases related to different initial positions of the PFOA molecule deposited on the 2D-Al QC. For each one of these cases, the PFOA molecule was initially geometrically optimized (see Figure 6). We then ran the AIMD simulations, monitored the functional groups and counted, for each one, the number of broken/unbroken bonds formed by the carbon and oxygen atoms bonded to fluorine ones. The corresponding movies of the simulations are available in the supplementary materials.

To facilitate the analysis, we show in **Figure 5d** the PFOA molecule indicating its molecular groups where we monitored the possible bond breaks of C-F (or C-O). Note that at the ends of the molecule, we have the $CF_3$ and $COO^-$ groups, while throughout the body chain, we have the different $CF_2$ groups, named from A to F.

In **Figure 5e**, we present the results of our analyses considering the different configurations. To count the number of unbroken bonds remaining after the PFOA interaction with the 2D-Al QC crystal, we analysed the remaining unbroken bonds for each group for the six simulations. In this way, we recorded the percentage of C-F or C-O bonds that remain unbroken after the interaction.

From these results, we verified that the bonds between the carbon and oxygen atoms of the $COO^-$ group are not broken after the interaction occurs. Therefore, these groups tend to remain in the molecule, which is consistent with the experimental results obtained by Raman, as shown in **Figure 5b**. In only one of the six simulations we observed that the bond between the carbon atom of the COO⁻ group and the carbon atom of the $CF_2^F$ group is broken. This indicates that even if the C-C of these atoms dissociate, the COO⁻ modes are observed due to the strong nature of their interaction. While we observed that the C-O bonds remain intact, we should also observe that the bonds of the $CF_3$ group should not break. Only the groups in the body of the PFOA should have the C-F bonds dissociated. However, we observed that both the C-F bonds of the $CF_3$ and $CF_2$ groups are broken after the interaction occurs. We observed in the simulations that, as except for the $CF_2^D$ group, where 84% of the C-F bonds remain unbroken, in all other cases, the number of unbroken C-F bonds is higher for the $CF_3$ group compared to the $CF_2$ groups. Based on our AIMD simulation results, we see that statistically, on average, a C-F bond in the $CF_3$ group is more likely to remain unbroken compared to those in the $CF_2$ group.

If we closely examine the experimental Raman results shown in **Figure 5c**, we see that for wavenumbers below 800 cm⁻¹, a small peak appears, indicating that some C-F modes belonging to the $CF_2$ groups are observed. Since this peak is much smaller than the one observed for the C-F modes of the $CF_3$ group, the simulation results are consistent with the experimental findings. The fact is that the number of C-F bond breakages in the $CF_2$ groups is much higher than in the $CF_3$ group. Therefore, for the concentration of PFOA considered in the experiment, the Raman shows a higher number of CF modes from the $CF_3$ groups, resulting in a higher intensity compared to the $CF_2$ group. In this sense, the interpretation of the experimental results discussed in this paper are supported and can be validated by what was observed in the AIMD simulations.

We have performed a comprehensive comparison with the contemporary studies on PFAS sensing and have enlisted the obtained detection limit with the methods of detection and materials used in **Table S2.**

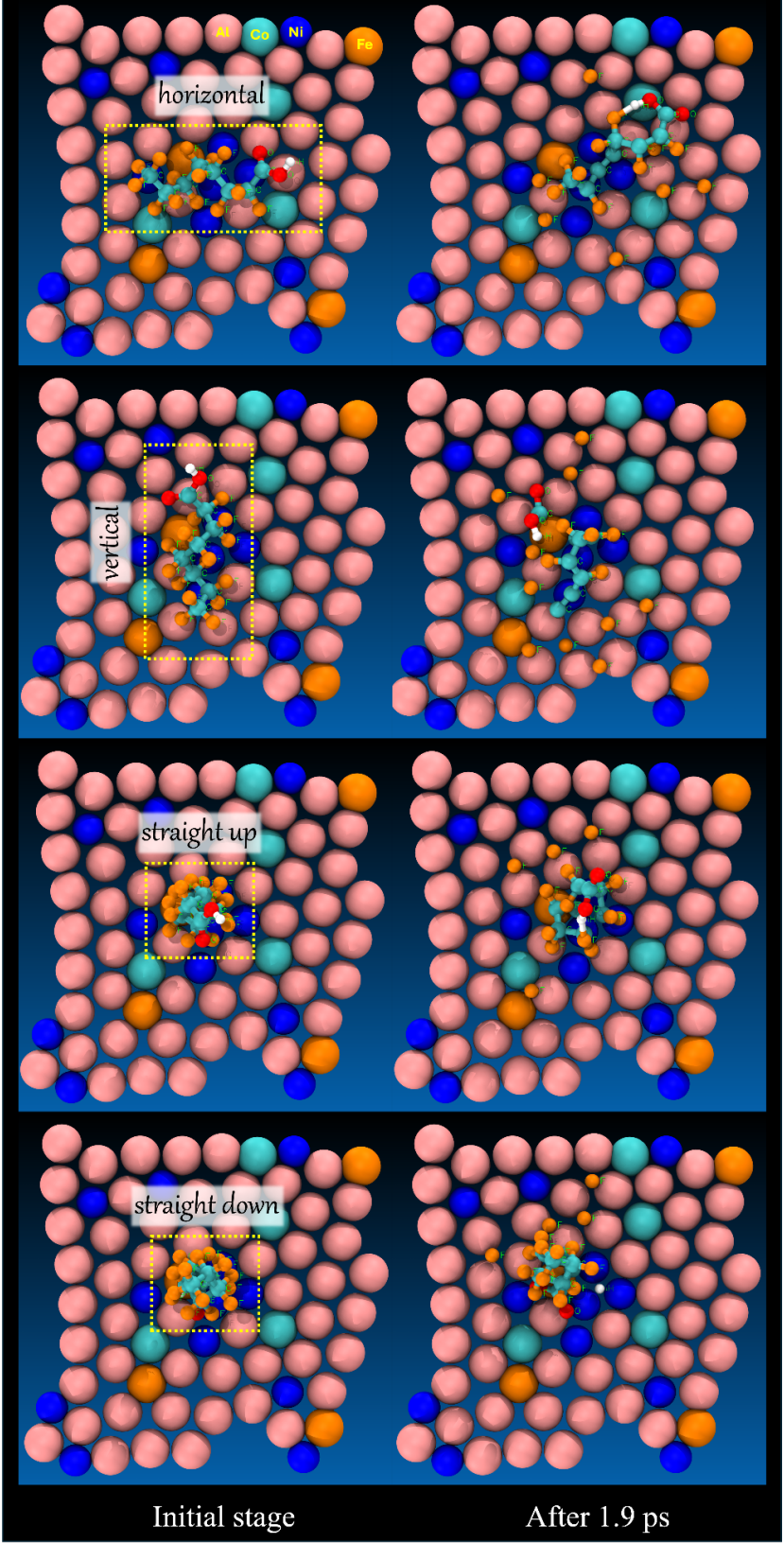

**Figure 6**. *Representative snapshots of the ab initio molecular dynamics (AIMD) simulations considering the system in a fixed volume at room temperature (300 K) for 1.9 ps. We considered six different configurational positions for the PFOA molecule.*

**Conclusion**

In this study, we successfully synthesized 2D-Al QC with a facile and highly scalable liquid exfoliation technique and demonstrated their potential for the subpicomolar level detection of PFOA. The 2D-Al QC-inked electrode shows a promising result with 0.59 pM LoD, high selectivity, excellent repeatability, and stability upto 90 days with a minor variation of 15%. The material's unique properties, such as high electron mobility, unique surface chemistry, large surface area, and abundant active sites, significantly enhanced its interaction with PFOA, enabling rapid and precise detection. Additionally, spectroscopic analysis and theoretical calculations provided valuable insights into the sensor-analyte interaction, confirming the strong binding affinity of the 2D-Al QC for PFOA. This work not only highlights the versatility of 2D quasicrystals in electrochemical sensing but also offers a promising approach for the development of efficient, portable, and environmentally sustainable sensors for PFAS monitoring. Further studies can expand the application of these materials to other pollutants, paving the way for more effective environmental remediation strategies.

**References**

(1) Garg, S.; Kumar, P.; Greene, G. W.; Mishra, V.; Avisar, D.; Sharma, R. S.; Dumée, L. F. Nano-Enabled Sensing of per-/Poly-Fluoroalkyl Substances (PFAS) from Aqueous Systems – A Review. *J. Environ. Manage.* **2022**, *308*, 114655. https://doi.org/10.1016/j.jenvman.2022.114655.

(2) Crone, B. C.; Speth, T. F.; Wahman, D. G.; Smith, S. J.; Abulikemu, G.; Kleiner, E. J.; Pressman, J. G. Occurrence of Per- and Polyfluoroalkyl Substances (PFAS) in Source Water and Their Treatment in Drinking Water. *Crit. Rev. Environ. Sci. Technol.* **2019**,

*49* (24), 2359–2396. https://doi.org/10.1080/10643389.2019.1614848.

(3) Gaines, L. G. T. Historical and Current Usage of Per- and Polyfluoroalkyl Substances (PFAS): A Literature Review. *Am. J. Ind. Med.* **2023**, *66* (5), 353–378. https://doi.org/10.1002/ajim.23362.

(4) Ding, N.; Harlow, S. D.; Randolph Jr, J. F.; Loch-Caruso, R.; Park, S. K. Perfluoroalkyl and Polyfluoroalkyl Substances (PFAS) and Their Effects on the Ovary. *Hum. Reprod. Update* **2020**, *26* (5), 724–752. https://doi.org/10.1093/humupd/dmaa018.

(5) Concellón, A.; Castro-Esteban, J.; Swager, T. M. Ultratrace PFAS Detection Using Amplifying Fluorescent Polymers. *J. Am. Chem. Soc.* **2023**, *145* (20), 11420–11430. https://doi.org/10.1021/jacs.3c03125.

(6) Thompson, D.; Zolfigol, N.; Xia, Z.; Lei, Y. Recent Progress in Per- and Polyfluoroalkyl Substances (PFAS) Sensing: A Critical Mini-Review. *Sensors and Actuators Reports* **2024**, *7*, 100189. https://doi.org/10.1016/j.snr.2024.100189.

(7) Fang, C.; Wu, J.; Sobhani, Z.; Amin, M. Al; Tang, Y. Aggregated-Fluorescent Detection of PFAS with a Simple Chip. *Anal. Methods* **2019**, *11* (2), 163–170. https://doi.org/10.1039/C8AY02382D.

(8) Chen, B.; Yang, Z.; Qu, X.; Zheng, S.; Yin, D.; Fu, H. Screening and Discrimination of Perfluoroalkyl Substances in Aqueous Solution Using a Luminescent Metal–Organic Framework Sensor Array. *ACS Appl. Mater. Interfaces* **2021**, *13* (40), 47706–47716. https://doi.org/10.1021/acsami.1c15528.

(9) Park, J.; Yang, K.-A.; Choi, Y.; Choe, J. K. Novel SsDNA Aptamer-Based Fluorescence Sensor for Perfluorooctanoic Acid Detection in Water. *Environ. Int.* **2022**, *158*, 107000. https://doi.org/10.1016/j.envint.2021.107000.

(10) Harrison, E. E.; Waters, M. L. Detection and Differentiation of Per- and

Polyfluoroalkyl Substances (PFAS) in Water Using a Fluorescent Imprint-and-Report Sensor Array. *Chem. Sci.* **2023**, *14* (4), 928–936. https://doi.org/10.1039/D2SC05685B.

(11) Trinh, V.; Malloy, C. S.; Durkin, T. J.; Gadh, A.; Savagatrup, S. Detection of PFAS and Fluorinated Surfactants Using Differential Behaviors at Interfaces of Complex Droplets. *ACS Sensors* **2022**, *7* (5), 1514–1523. https://doi.org/10.1021/acssensors.2c00257.

(12) Concellón, A.; Swager, T. M. Detection of Per- and Polyfluoroalkyl Substances (PFAS) by Interrupted Energy Transfer. *Angew. Chemie Int. Ed.* **2023**, *62* (47). https://doi.org/10.1002/anie.202309928.

(13) Calvillo Solís, J. J.; Yin, S.; Galicia, M.; Ersan, M. S.; Westerhoff, P.; Villagrán, D. "Forever Chemicals" Detection: A Selective Nano-Enabled Electrochemical Sensing Approach for Perfluorooctanoic Acid (PFOA). *Chem. Eng. J.* **2024**, *491*, 151821. https://doi.org/10.1016/j.cej.2024.151821.

(14) Clark, R. B.; Dick, J. E. Electrochemical Sensing of Perfluorooctanesulfonate (PFOS) Using Ambient Oxygen in River Water. *ACS Sensors* **2020**, *5* (11), 3591–3598. https://doi.org/10.1021/acssensors.0c01894.

(15) Kazemi, R.; Potts, E. I.; Dick, J. E. Quantifying Interferent Effects on Molecularly Imprinted Polymer Sensors for Per- and Polyfluoroalkyl Substances (PFAS). *Anal. Chem.* **2020**, *92* (15), 10597–10605. https://doi.org/10.1021/acs.analchem.0c01565.

(16) Karimian, N.; Stortini, A. M.; Moretto, L. M.; Costantino, C.; Bogialli, S.; Ugo, P. Electrochemosensor for Trace Analysis of Perfluorooctanesulfonate in Water Based on a Molecularly Imprinted Poly( o -Phenylenediamine) Polymer. *ACS Sensors* **2018**, *3* (7), 1291–1298. https://doi.org/10.1021/acssensors.8b00154.

(17) Glasscott, M. W.; Vannoy, K. J.; Kazemi, R.; Verber, M. D.; Dick, J. E. μ-MIP:

Molecularly Imprinted Polymer-Modified Microelectrodes for the Ultrasensitive Quantification of GenX (HFPO-DA) in River Water. *Environ. Sci. Technol. Lett.* **2020**, *7* (7), 489–495. https://doi.org/10.1021/acs.estlett.0c00341.

(18) Khan, R.; Andreescu, D.; Hassan, M. H.; Ye, J.; Andreescu, S. Nanoelectrochemistry Reveals Selective Interactions of Perfluoroalkyl Substances (PFASs) with Silver Nanoparticles. *Angew. Chemie Int. Ed.* **2022**, *61* (42). https://doi.org/10.1002/anie.202209164.

(19) Khan, R.; Uygun, Z. O.; Andreescu, D.; Andreescu, S. Sensitive Detection of Perfluoroalkyl Substances Using MXene–AgNP-Based Electrochemical Sensors. *ACS Sensors* **2024**, *9* (6), 3403–3412. https://doi.org/10.1021/acssensors.4c00776.

(20) Mukhopadhyay, N. K.; Yadav, T. P. Quasicrystals: A New Class of Structurally Complex Intermetallics. *J. Indian Inst. Sci.* **2022**, *102* (1), 59–90. https://doi.org/10.1007/s41745-022-00293-1.

(21) Yadav, T. P.; Woellner, C. F.; Sinha, S. K.; Sharifi, T.; Apte, A.; Mukhopadhyay, N. K.; Srivastava, O. N.; Vajtai, R.; Galvao, D. S.; Tiwary, C. S.; Ajayan, P. M. Liquid Exfoliation of Icosahedral Quasicrystals. *Adv. Funct. Mater.* **2018**, *28* (26). https://doi.org/10.1002/adfm.201801181.

(22) Mishra, S. S.; Kumbhakar, P.; Nellaiappan, S.; Katiyar, N. K.; Tromer, R.; Woellner, C. F.; Galvao, D. S.; Tiwary, C. S.; Ghosh, C.; Dasgupta, A.; Biswas, K. Two-Dimensional Multicomponent Quasicrystal as Bifunctional Electrocatalysts for Alkaline Oxygen and Hydrogen Evolution Reactions. *Energy Technol.* **2023**, *11* (2). https://doi.org/10.1002/ente.202200860.

(23) Yadav, T. P.; Woellner, C. F.; Sharifi, T.; Sinha, S. K.; Qu, L.; Apte, A.; Mukhopadhyay, N. K.; Srivastava, O. N.; Vajtai, R.; Galvão, D. S.; Tiwary, C. S.; Ajayan, P. M. Extraction of Two-Dimensional Aluminum Alloys from Decagonal

Quasicrystals. *ACS Nano* **2020**, *14* (6), 7435–7443. https://doi.org/10.1021/acsnano.0c03081.

(24) Kumar, S.; Chaurasiya, R.; Mishra, S. S.; Kumbhakar, P.; Meng, G.; Tiwary, C. S.; Biswas, K.; Kumar, M. Nanocomposites of Quasicrystal Nanosheets and MoS 2 Nanoflakes for NO 2 Gas Sensors. *ACS Appl. Nano Mater.* **2023**, *6* (7), 5952–5962. https://doi.org/10.1021/acsanm.3c00346.

(25) Kumar, S.; Hojamberdiev, M.; Chakraborty, A.; Mitra, R.; Chaurasiya, R.; Kwoka, M.; Tiwary, C. S.; Biswas, K.; Kumar, M. Quasicrystal Nanosheet/α-Fe 2 O 3 Heterostructure-Based Low Power NO 2 Sensors: Experimental and DFT Studies. *ACS Appl. Mater. Interfaces* **2024**, *16* (13), 16687–16698. https://doi.org/10.1021/acsami.4c00201.

(26) Mandal, N.; Kumbhakar, P.; Dey, A.; Kumbhakar, P.; Chatterjee, U.; J. S. de Matos, C.; Prasad Yadav, T.; Krishna Mukhopadhyay, N.; Biswas, K.; Kochat, V.; Sekhar Tiwary, C. Optical Resonator-Enhanced Random Lasing Using Atomically Thin Aluminium-Based Multicomponent Quasicrystals. *Opt. Laser Technol.* **2024**, *175*, 110746. https://doi.org/10.1016/j.optlastec.2024.110746.

(27) Kumbhakar, P.; Pramanik, A.; Mishra, S. S.; Tromer, R.; Biswas, K.; Dasgupta, A.; Galvao, D. S.; Tiwary, C. S. Enhanced Light Scattering Using a Two-Dimensional Quasicrystal-Decorated 3D-Printed Nature-Inspired Bio-Photonic Architecture. *J. Phys. Chem. C* **2023**, *127* (20), 9779–9786. https://doi.org/10.1021/acs.jpcc.3c00513.

(28) Kumar, S.; Chaurasiya, R.; Mishra, S. S.; Kumbhakar, P.; Meng, G.; Tiwary, C. S.; Biswas, K.; Kumar, M. Nanocomposites of Quasicrystal Nanosheets and MoS2 Nanoflakes for NO2 Gas Sensors. *ACS Appl. Nano Mater.* **2023**, *6* (7), 5952–5962. https://doi.org/10.1021/acsanm.3c00346.

(29) Ramachandran, R.; Leng, X.; Zhao, C.; Xu, Z. X.; Wang, F. 2D Siloxene Sheets: A

Novel Electrochemical Sensor for Selective Dopamine Detection. *Appl. Mater. Today* **2020**, *18*, 100477. https://doi.org/10.1016/j.apmt.2019.100477.

(30) Sukanya, R.; Sakthivel, M.; Chen, S. M.; Chen, T. W. A New Type of Terbium Diselenide Nano Octagon Integrated Oxidized Carbon Nanofiber: An Efficient Electrode Material for Electrochemical Detection of Morin in the Food Sample. *Sensors Actuators, B Chem.* **2018**, *269*, 354–367. https://doi.org/10.1016/j.snb.2018.05.013.

(31) Murugan, N.; Jerome, R.; Preethika, M.; Sundaramurthy, A.; Sundramoorthy, A. K. 2D-Titanium Carbide (MXene) Based Selective Electrochemical Sensor for Simultaneous Detection of Ascorbic Acid, Dopamine and Uric Acid. *J. Mater. Sci. Technol.* **2021**, *72*, 122–131. https://doi.org/10.1016/j.jmst.2020.07.037.

(32) Mahapatra, P. L.; Campos de Oliveira, C.; Sreeram, P. R.; Sivaraman, S. K.; Sarkar, S.; Costin, G.; Lahiri, B.; Autreto, P. A. da S.; Tiwary, C. S. Hydrogen Sulfide Gas Detection Using Two-Dimensional Rhodonite Silicate. *Chem. Mater.* **2023**, *35* (19), 8135–8144. https://doi.org/10.1021/acs.chemmater.3c01593.

(33) Liu, Q.; Huang, A.; Wang, N.; Zheng, G.; Zhu, L. Rapid Fluorometric Determination of Perfluorooctanoic Acid by Its Quenching Effect on the Fluorescence of Quantum Dots. *J. Lumin.* **2015**, *161*, 374–381. https://doi.org/10.1016/j.jlumin.2015.01.045.

(34) Yan, W.; Qian, T.; Zhang, L.; Wang, L.; Zhou, Y. Interaction of Perfluorooctanoic Acid with Extracellular Polymeric Substances - Role of Protein. *J. Hazard. Mater.* **2021**, *401*, 123381. https://doi.org/10.1016/j.jhazmat.2020.123381.

(35) Gong, Y.; Wang, L.; Liu, J.; Tang, J.; Zhao, D. Removal of Aqueous Perfluorooctanoic Acid (PFOA) Using Starch-Stabilized Magnetite Nanoparticles. *Sci. Total Environ.* **2016**, *562*, 191–200. https://doi.org/10.1016/j.scitotenv.2016.03.100.

(36) Liu, Y.; Hu, X.-M.; Zhao, Y.; Wang, J.; Lu, M.-X.; Peng, F.-H.; Bao, J. Removal of

Perfluorooctanoic Acid in Simulated and Natural Waters with Different Electrode Materials by Electrocoagulation. *Chemosphere* **2018**, *201*, 303–309. https://doi.org/10.1016/j.chemosphere.2018.02.129.

(37) Chen, Y.; Yang, Y.; Cui, J.; Zhang, H.; Zhao, Y. Decoding PFAS Contamination via Raman Spectroscopy: A Combined DFT and Machine Learning Investigation. *J. Hazard. Mater.* **2024**, *465*, 133260. https://doi.org/10.1016/j.jhazmat.2023.133260.